\documentclass[engish]{article}

\usepackage{hyperref}
\usepackage{cleveref}
\usepackage{color}
\usepackage[usenames,dvipsnames,svgnames,table]{xcolor}
\usepackage[english]{babel}
\usepackage{graphicx}
\usepackage{natbib}
\usepackage{ulem}

\newcommand{\aap}{Astron.~Astrophys.}  
\newcommand{\apjl}{Astrophys.~J.~Letters}   
\newcommand{\apjs}{Astrophys.~J.~Supp.}   

\small

\begin{document}

\twocolumn

\begin{center}
\fboxrule0.02cm
\fboxsep0.4cm
\fcolorbox{blue}{AliceBlue}{\rule[-0.9cm]{0.0cm}{1.8cm}{\parbox{7.8cm}
{ \begin{center}
{\Large\em Perspective}

\vspace{0.2cm}

{\large\bf Supersize me! Jets from massive young stellar objects} 



\vspace{0.2cm}

{\large\em Alessio Caratti o Garatti}

\vspace{0.5cm}

\centering
\includegraphics[width=0.24\textwidth]{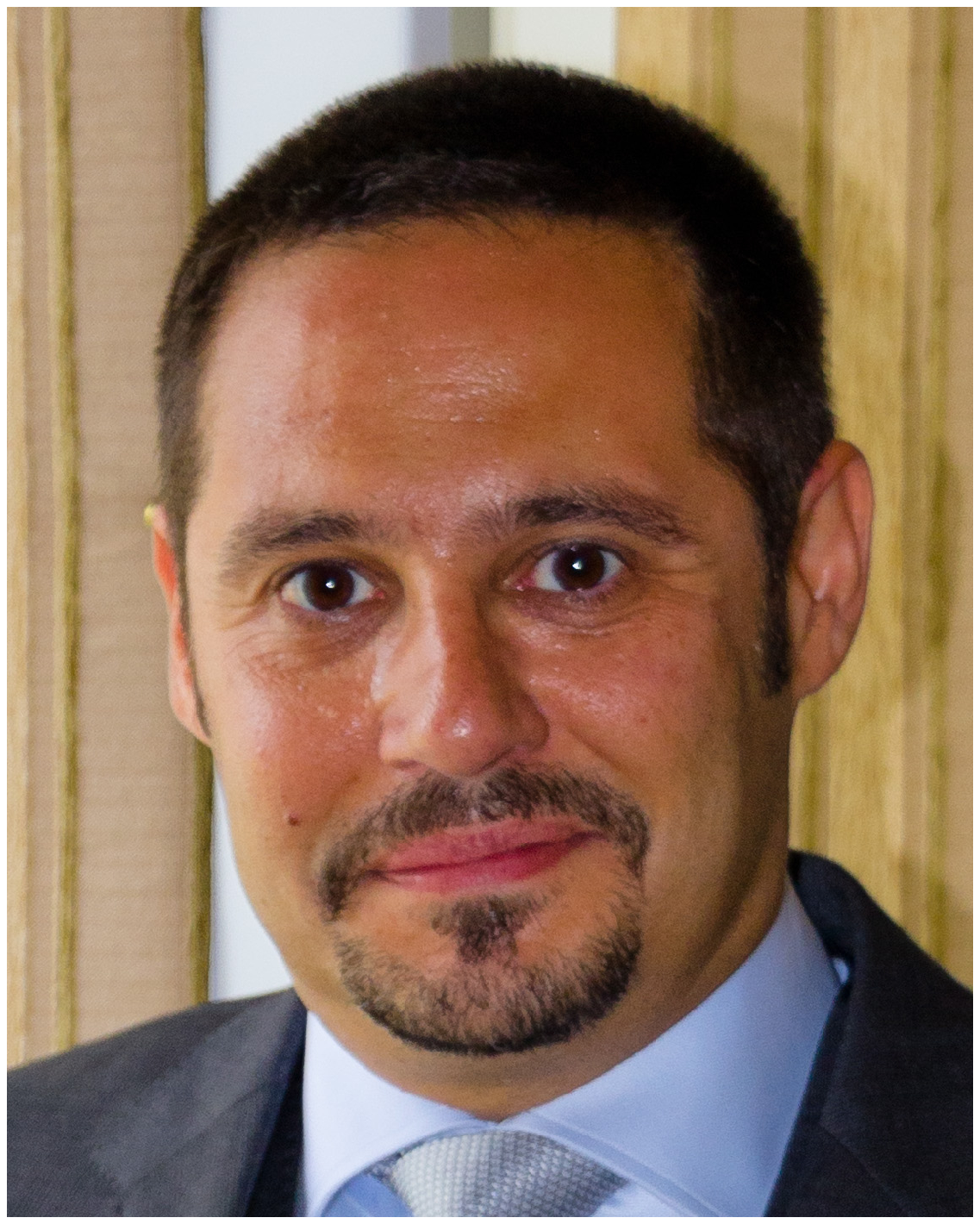}
\end{center}
}}}
\end{center}

\hyphenation{a-na-ly-sis mo-le-cu-lar pre-vious e-vi-den-ce di-ffe-rent pa-ra-me-ters ex-ten-ding a-vai-la-ble ca-li-bra-tion magne-to cen-tri-fu-gal in-ter-me-dia-te con-ti-nu-um spectro-in-ter-fe-ro-me-try spectro-as-tro-me-try mas-ses}

\subsection*{Introduction}
Jets and outflows (here the latter are identified as swept-up circumstallar/interstellar material) are a common outcome of the star formation process 
(from proto-brown-dwarfs to high-mass young stellar objects - HMYSOs; $M_*$ $\ge$ 8\,M$_\odot$, $L_{bol}$ $\ge$ 5$\times$10$^3$\,L$_\odot$) and are intimately linked to accretion disks and the YSO accretion
mechanism~\citep[for a review see][]{reipurth01,bally16}.
Although the jet launching mechanism is still debated (disk-wind, X-wind, stellar wind, etc..), there is growing agreement that jets are magnetically driven and collimated (this is certainly true 
for low-mass YSOs). Jets are also an essential ingredient for removing angular momentum in excess and allowing accretion to occur. Moreover, they play an important role in dispersing the 
protostellar envelope and, as a consequence, in regulating the final stellar mass. 

Our current understanding of jet formation and propagation mostly derives from studies of jets driven by low- and intermediate-mass YSOs. 
Indeed, due to the relatively small distance to us ($\geq$120\,pc)
and typical low visual extinction (a few A$_V$) towards low-mass star forming regions (SFRs), jet dynamical and physical parameters 
can be easily determined along the whole length of the jets~\citep[for a review see e.\,g.][]{ray07,frank14}. 
In addition, optical/NIR observations at high-angular and spectral resolution are providing us with a complete picture of the jet launching region and mechanisms, through spectro-interferometry~\citep[e.\,g.][]{rebeca15,rebeca16,rebeca17,caratti15b},
as well as the magnetic configuration and geometry of the YSO driving the flow, through spectro-polarimetry~\citep[e.\,g.][]{johns-krull09,johns-krull13}. 

On the other hand, the study of jets from HMYSOs is still in its infancy and their analysis is mostly limited by large distances (several kpc) and very high visual extinction (A$_V$ up to 100 mag) 
towards massive SFRs. Protostellar jets produce shocks that can be mainly studied at UV, optical and IR wavelengths. Therefore the high A$_V$ 
makes the detection and study of high-mass protostellar jets extremely challenging compared to their outflows, well studied through molecular
tracers (e.\,g. SiO, CO, HCO+, and their isotopologues) at submillimetre (sub-mm) and millimetre (mm) wavelengths~\citep[e.\,g.][]{beuther02,lopez-sepulcre09,lopez-sepulcre11,duarte,sanna14}.

\begin{figure}[htb]
\begin{center}
\includegraphics[width=0.5\textwidth]{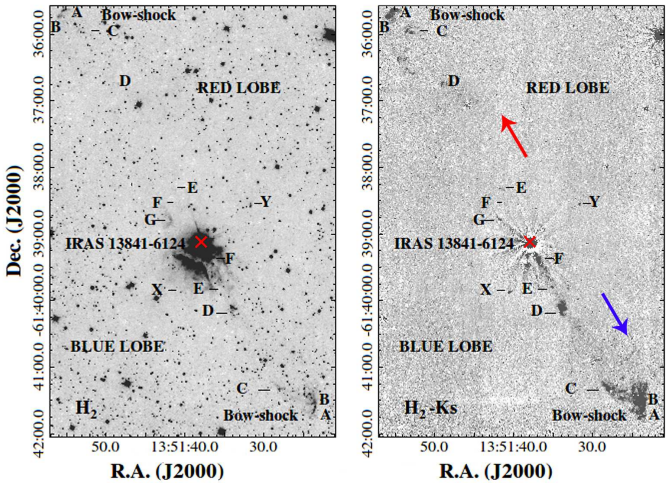}
\caption{H$_2$ and continuum-subtracted H$_2$ images (left and right panels) of the IRAS\,13481-6124 parsec-scale jet~\citep[from][]{caratti15a}. 
This is one of the few examples of massive jets with
small precession angles. Source and knots positions are indicated in the figures.}
\label{fig1:fig}
\end{center}
\end{figure}

Until recently, only a few massive jets were known~\citep[partially detected at optical wavelengths, 
e.\,g. HH\,80/81, HH135/136, HH\,168, HH\,396/397; e.\,g.,][]{reipurth01}. In the past decade, thanks to narrow-band (H$_2$ at 2.12\,$\mu$m and [FeII] at 1.64\,$\mu$m) near-infrared 
(NIR) imaging surveys~\citep[e.\,g.][]{varricatt10,stecklum09,froebrich11,lee12,caratti15a,navarete15,wolf-chase17}, it has become clear that also HMYSOs (at least up to 30-40\,M$_\odot$, e.\,g. late O-type stars) 
drive bipolar collimated jets (see Fig.~\ref{fig1:fig}). These massive jets most likely originate from massive accretion disks/tori, which have been detected in several HMYSOs with masses 
up to 60\,M$_\odot$~\citep[e.\,g.][]{ilee16}. The lack of jets driven by more massive YSOs could be due to: \textit{i)} an observational bias, as there are just a few early O-type HMYSOs 
and these are highly embedded; \textit{ii)} the ejection mechanism is not the same for the most massive protostars. Most recent magneto-hydrodynamic (MHD; Vaidya et al. 2011) and hydrodynamic~\citep[][]{kuiper15,kuiper16}
simulations seem to indicate that stellar radiation plays a major role in affecting the jet dynamics. As the stellar mass (and thus its radiation) increases with age,
jets/outflows become less and less collimated~\citep[e.\,g.][]{beuther05}. 

More generally, the powerful outflows driven by massive jets can be well 
or poorly collimated~\citep[e.\,g.][]{arce07}. The poor collimation can be caused by different factors: a) highly precessing jets; b) jet multiplicity; c) wide-angle winds with or without jets; d) stellar radiation;
e) explosive events. In this short review, I will not discuss explosive outflows~\citep[as, e.\,g., Orion KL and DR\,21; see e.\,g.][]{bally16}, which are likely associated with the disruption 
of a non-hierarchical massive and young stellar system, and triggered by the possible merger of HMYSOs or by protostellar collision~\citep[e.\,g.][]{bally16,zapata17}. 

The past few years have seen additional breakthroughs in the field of massive jets.
Despite our poor knowledge of magnetic fields in YSOs, magnetic fields have been detected in massive jets and close to HMYSOs, with geometries
parallel to the jet axis and perpendicular to the disk~\citep[][]{carrasco10,sanna15}. This supports an MHD origin and thus suggests that the accretion/ejection processes 
in HMYSOs could be somehow similar to those observed in low-mass YSOs, despite the very different environmental conditions. 
Further evidence supporting the MHD origin of massive jets comes from the 
observational signature of a rotating outflow driven by a magneto centrifugal disk-wind launched by Orion Source I in the KL region~\citep[][]{hirota17}.
 
Finally, protostellar jets are an indirect tracer of the accretion process in YSOs, providing us with fundamental clues
about the accretion processes and the accretion history of YSOs~\citep[e.\,g.][]{arce07,bally07}. In the case of embedded YSOs, jets become particularly important, as
the accretion process cannot be directly observed. At variance with outflows, jets have the advantage to provide us with the direct 
measurement of YSO mass loss rates at different epochs and thus, indirectly, with an estimate of accretion rates. The interconnection between accretion and ejection 
activities in HMYSOs has been finally proven by the recent observations of a radio jet outburst in S255 NIRS\,3~\citep[][]{cesaroni18}, which followed an accretion outburst of this 20\,M$_\odot$ HMYSO~\citep[][]{caratti17}.
The time lag (about one year) between the IR accretion outburst and the radio jet burst indicates that the latter must have been triggered mechanically. 
This proves that the radio jet has been boosted by a sudden increase of the mass loss rate.

\subsection*{IR jet tracers}
Protostellar jets are launched at supersonic velocity, producing shocks observed at UV, optical and IR wavelengths, 
where the brightest ionic (e.\,g. [O I], [S II], HI, [FeII]) and molecular (i.e. H$_2$) jet tracers occur~\citep{reipurth01}. 
Given the high visual extinctions, the UV and optical regimes are usually not accessible, thus we have to rely on infrared tracers to study massive jets. 
In the near- and mid-IR regimes, molecular hydrogen is the main jet tracer and coolant~\citep[][]{caratti15a}. For instance, the bright 1-0\,S(1) line at 2.12\,$\mu$m and many 
other ro-vibrational H$_2$ lines trace temperatures from a thousand to several thousands K whereas the pure-rotational lines in the mid-IR trace temperatures from several hundreds to a thousand K~\citep[see e.\,g.][]{caratti08,caratti15a}. 
The H$_2$ emission typically comes from shocks in jets with bow-shock-, knot- or jet-like morphologies. However, it can also have a fluorescent origin 
from photodissociation regions (PDRs) associated to UCHII/HII regions (close to the driving HMYSO) or excited by OB stars (with a bubble-like or filament-like morphology)~\citep[see][]{wolf-chase17}. 
The nature of the H$_2$ excitation can be established through line ratios (e.\,g. 1-0\,S(1)/2-1\,S(1), i.e. 2.12/2.25\,$\mu$m)~\citep[see][]{wolf-chase17}.
Other relevant NIR lines, so far detected along massive jets, are those from the [FeII] (with its brightest lines at 1.257 and 1.644\,$\mu$m, the latter detected in $\sim$50\% of flows)~\citep[][]{caratti15a}
and HI species (observed less frequently, from the Paschen and Brackett series). Both species trace strong dissociative shocks. 
HI emission has also been detected spectroscopically close or coincident to the HMYSO continuum. 
In this case HI emission could also be pumped by the strong UV radiation field of HMYSOs. However, there are a few clear cases where the Br$\gamma$ line is clearly emitted at the base of the
jet (see Sect. ``Jets close to source''). 

The MIR and FIR wavelength regimes have been less explored, as they are mostly unobservable from ground. The most important and promising tracer of massive jets is the [OI] lines
at 63 and 145\,$\mu$m (the 63\,$\mu$m line being one of the most important atomic coolants in jets). These lines are
crucial for studying the flow in the most embedded HMYSOs~\citep[see e.\,g.][]{leurini15}.   
An unsolved issue in the NIR/MIR wavelength regime is represented by the so-called extended green objects (EGOs)~\citep[][]{cyganowski08}, 
observed by the Infrared Array Camera (IRAC) on the Spitzer Space Telescope. EGOs show emission in excess in band 2 (at 4.5\,$\mu$m) and their name comes
from the common colour-coding of IRAC band 2 in the three-colour composite images (at 3.6, 4.5, and 8\,$\mu$m). EGOs have been detected along the jets as well as around HMYSOs. 
While in the former case such emission likely comes from shocks~\citep[H$_2$ pure-rotational lines and CO fundamental vibrational bands][]{tappe12},
the latter case is less clear as emission might come from the HMYSO continuum, scattered by the outflow cavities~\citep[][]{takami12}. 

\subsection*{Morphology, Physics and Kinematics}

The morphology of massive jets is quite varied, ranging from simple straight bipolar geometries (representing, so far, the minority of the objects; see e.\,g. Figure~\ref{fig1:fig}) to more complex
ones, usually highly precessing and even monopolar.
Indeed HMYSOs usually form in clusters or small associations. Therefore N-body interactions must be a fundamental ingredient in shaping the jets, producing changes in the flow
orientation and thus large precession angles. Although precession is also observed in low-mass YSOs, HMYSOs tend to have large precession angles (up to 40$^{\circ}$--50$^{\circ}$) as well as
abrupt changes in the jet direction, which strongly affect the outflow collimation. Jet multiplicity is another characteristic of massive SFRs, originating from the high stellar multiplicity 
and large distances to massive SFRs. NIR imaging surveys show that some of these massive jets are monopolar~\citep[e.\,g. 39\% of the sample in][]{caratti15a}.
While the high visual extinction can
possibly explain the non-detection of the red-shifted lobe for several objects, it is also possible that other mechanisms, which inhibit the bipolar ejection, are at play
(e.\,g. anisotropic ambient cloud conditions or the presence of close companions; see e.\,g. Murphy et al. 2008). Notably, also some outflows from low-mass~\citep[][]{codella14} 
and high-mass YSOs~\citep[][]{zapata06,fernandez-lopez} are known to be monopolar. Despite the different observed morphologies, jets are well collimated, extending up to parsec scales. 
Moreover, the presence of both jet-like and knot-like structures might
indicate that matter flows continuously with episodic enhancements, which produce knotty structures along the flows~\citep[][]{caratti17,cesaroni18}. 

Despite the increasing number of detected massive jets through NIR imaging surveys, just a small number have been studied spectroscopically in detail at 
NIR wavelengths \citep[][]{davis04,gredel06,caratti08,caratti15a,fedriani18}. Accordingly, our knowledge of their physical and kinematic properties 
is still quite limited. 

\begin{figure}[htb]
\begin{center}
\includegraphics[width=0.5\textwidth]{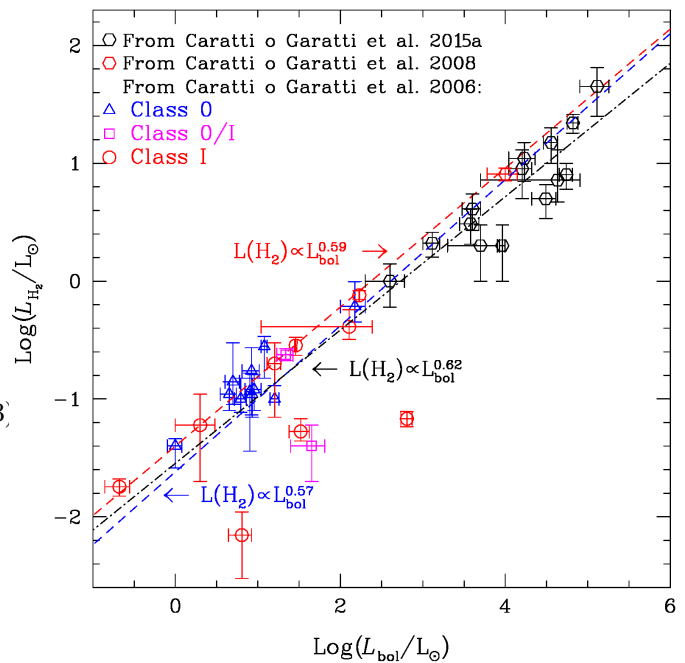}
\caption{$Log\,(L_{H2})\,vs.\,Log\,(L_{bol})$ for low-mass and high-mass jets~\citep[adapted from][]{caratti15a}. The red, black and blue dashed lines indicate the best linear
fit resulting from the low-mass sample, high-mass sample and combined samples, respectively.}
\label{fig2:fig}
\end{center}
\end{figure}

NIR spectroscopy has proven to be a powerful tool to derive the main physical and dynamical properties of massive jets.
For example, ro-vibrational diagrams of the H$_2$ shocked emission can provide the main physical parameters
along the jet~\citep[i.\,e. temperature, column density, visual extinction; see][]{davis04,caratti08,caratti15a}. 
This type of analysis allows us to estimate the H$_2$ luminosity ($L_{H2}$) from HMYSO jets, which should represent a significant fraction of the overall energy radiated away in the cooling process.
$L_{H2}$ and the bolometric luminosity of the driving source appear to be related ($L_{H2} \sim 0.03 \times L_{bol}^{0.6}$; see Fig.~\ref{fig2:fig}), correlation which holds from low-mass to HMYSOs~\citep[][]{caratti06,caratti08,caratti15a}.
This relation closely resembles those found for the mechanical force of the CO ($\dot{P}_{CO}$) outflows~\citep[$\dot{P}_{CO} \propto L_{bol}^{0.6}$; see][]{beuther02,duarte,maud15}
as well as for the radio continuum luminosity of the jets~\citep[$L_{jet} \propto L_{bol}^{0.54}$; see e.\,g.][for a review]{anglada}.
These correlations directly link the outflow/jet properties with the driving source properties.
These results can be understood theoretically for sources that derive most of their luminosity from accretion. Notably, such correlations
extend from massive young stars to the sub-stellar domain, suggesting a common formation mechanism.

On average, the physical parameters of jets from HMYSOs are greater than those from their low-mass siblings, showing much higher visual extinction values (up to 50--100\,mag),
much higher column densities (up to 10$^{20}$\,cm$^{-2}$) and masses (up a few M$_{\odot}$), higher H$_2$ temperatures (2500--3000\,K) and larger luminosities (up to 50\,L$_{\odot}$ for
HMYSOs with $\sim$10$^5$\,L$_{bol}$). For what concerns the kinematic analysis, massive jets appear to be moving faster than their low-mass counterparts,
with velocities up to 800\,km/s in jets driven by a 20\,M$_{\odot}$ HMYSO~\citep[][]{fedriani18}. This is not unexpected, as the escape velocity of jets from massive protostars must be higher than 
that of the low-mass ones. Notably, HMYSO jet velocities have been estimated so far from radial velocities, knowing the jet/disk geometries, but jet precession angles are usually not taken into account.
Indeed a 3D kinematic study (which combines radial and tangential velocities) is still missing.

Estimates of mass-ejection rates seem to be much higher~\citep[i.\,e. from several 10$^{-6}$ to several 10$^{-4}$\,M$_{\odot}$\,yr$^{-1}$ for jets driven by HMYSOs in the range of 8--20\,M$_{\odot}$][]{caratti08,mcLeod18,fedriani18} 
than in low-mass jets, also suggesting that mass accretion rates are higher in HMYSOs, as already inferred through molecular tracers; 10$^{-4}$--10$^{-3}$\,M$_{\odot}$\,yr$^{-1}$~\citep[see e.\,g.][]{beltran16,beuther17}.
Also other dynamical properties, such as the mechanical force, kinetic energy, and momentum seem to scale with the jet mass (Caratti o Garatti et al. 2008; Fedriani et al. 2018).

As a matter of fact, massive jets appear to be scaled-up versions of their low-mass counterparts, in agreement with the idea that the launching mechanism is similar, if not the same. 
However, there are also notable differences, such as more diverse morphologies, likely due to their different environments, and, most importantly, the ionised nature of the jet
close to the source. Nevertheless jets from low-mass stars can be ionised as well, especially by external irradiation from nearby massive stars~\citep[see e.\,g.][]{bally16}.

\subsection*{Jets close to source}
A crucial difference between high- and low-mass YSOs resides in the fact that massive protostars reach the zero age main sequence while they are still accreting.
As a consequence, their hot photospheres become a strong source of Lyman radiation, which ionises and disrupts the HMYSO circumstellar environment. 
Although it is not yet completely clear how massive jets form, collimate and survive in such a ``hostile'' environment~\citep[but several theoretical models and simulations are able to explain and reproduce them,
see e\,g.][]{vaidya11,seifried13,kuiper15,kuiper16}, observations tell us that they exist. The presence of massive disks must thus play a fundamental role in shielding and launching them.
Additionally, an important role could be played by the high density at the base of the jet, which might trap the stellar ionising radiation, so that the jet beam is neutral at larger distances~\citep[][]{raga}. 
The structure of massive jets close to source might then resemble an onion-like structure, with an ionised base and core plus layers of neutral atomic, and (possibly) molecular gas.
In this hypothetical picture, also the jet/outflow velocity would show an onion-like structure, moving from the axis to the outer regions of the cone~\citep[][]{koelligan17}. 

\begin{figure}[htb]
\begin{center}
\includegraphics[width=0.5\textwidth]{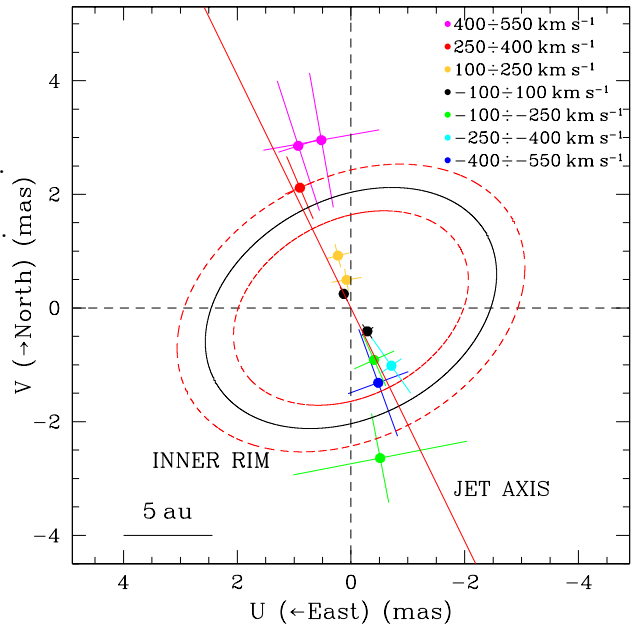}
\caption{Br$\gamma$ line displacement for different (radial) velocity channels (with
velocity bins from -550 to 550\,km\,s$^{−1}$) at the base of IRAS\,13481-6124 jet~\citep[from][]{caratti16}.}
\label{fig3:fig}
\end{center}
\end{figure}

An important piece of the puzzle is represented by the detection and study of radio jets close (or coincident) to HMYSOs, through high angular resolution radio continuum observations~\citep[][]{guzman12,moscadelli16,purser16},
showing elongated (often bipolar) structures tracing the fully ionised jet~\citep[see][for a review]{anglada}. 
Along with a fully ionised jet a highly collimated HI jet (in the NIR, traced by Br$\gamma$ and Br$\alpha$ lines) has been directly observed in IRAS\,13481-6124 (with spectro-interferometry 
and spectro-astrometry; see Fig.~\ref{fig3:fig}) \citep[][]{caratti16,fedriani18} and, indirectly in other HMYSOs, through the detection on source of P\,Cygni line profiles in the Brackett series~\citep[][]{cooper13}.
In most cases, the NIR jet emission cannot be detected due 
to the prohibitive visual extinction, thus our understanding of the launching region must rely
on the radio jet continuum \citep[][]{anglada} and, possibly, on hydrogen recombination lines recently detected with ALMA~\citep[][]{guzman14}. 
Their combined analysis can provide us with physical, kinematic and dynamical properties, such as velocities or ionised mass loss rates. The latter are found to be 
at least one order of magnitude smaller than mass ejection rates derived from molecular outflows~\citep[e.\,g.][]{guzman12} and jets~\citep[][]{fedriani18}, indicating that jets are only partially
ionised~\citep[$\sim$10\%; see e.\,g.][]{cesaroni18,fedriani18}.

\subsection*{Conclusions \& Future Perspective}

Massive jets are a common byproduct of massive star formation, detected towards a large number of HMYSOs. 
There is growing observational evidence that massive jets are a scaled-up version of their low-mass siblings (in terms of physics and kinematics). 
However, the studied sample is still too small to draw firm conclusions, therefore IR spectroscopic surveys (with e.\,g. JWST), on large samples
(spanning a wider range of HMYSO masses and evolutionary stages) are required. Detection and study of magnetic fields in a larger number of massive jets
is also envisaged. The next generation of radio interferometers, as LOFAR, SKA, ngVLA, will provide us with more clues regarding the non-thermal emission of massive jets.

The innermost region close to HMYSOs (within 100\,au from the source) is key for understanding 
how jets are launched and collimated, thus the deployment of high angular and spectral resolution is a fundamental step forward. 
NIR and MIR spectro-interferometry (e.\,g. with the ESO-VLTI second generation interferometers GRAVITY and MATISSE at milliarcsecond resolution) 
are powerful tools, complementary to mm and radio interferometry, to study the origin of massive jets.


{\bf Acknowledgements}

It has been a pleasure receiving comments and suggestions from Jochen Eisl\"offel, Tom Ray, Bringfried Stecklum, Rebeca Garcia Lopez and Anna McLeod.
I acknowledge support from the European Research Council (ERC) under the European Union's Horizon 2020 research and innovation programme (grant agreement No.\ 743029).



\bibliographystyle{aa}
\bibliography{references.bib}
\end{document}